\documentclass{iau} 
\usepackage{graphicx}

\title[Galaxy archeology with deep imaging] %% give here short title %%
{Using deep images and simulations to trace  collisional debris around massive galaxies}

\author[Pierre-Alain Duc]   %% give here short author list %%
{Pierre-Alain Duc$^1$
\affiliation{$^1$AIM Paris-Saclay \\ Service d'astrophysique, CEA-Saclay, 91191 Gif sur Yvette, France \\ email: {\tt paduc@cea.fr} \\[\affilskip]
}}

\pubyear{2016}
\volume{321}  %% insert here IAU Symposium No.
\jname{Formation and evolution of galaxy outskirts}
\editors{Gil de Paz, A.}
\begin{document}

\maketitle

\begin{abstract}
Deep imaging programs, such as MATLAS which  has just been completed at  the CFHT,  allows us to study with their diffuse light the outer stellar populations  around  large number of galaxies. We have carried out a systematic census of their  fine structures, i.e. the collisional debris from past mergers. We have identified among them  stellar streams from minor  mergers, tidal tails from gas-rich major mergers, plumes from gas-poor major mergers, and shells from intermediate mass mergers.  Having estimated the visibility and life time of each of these structures with numerical simulations, we can reconstruct the past mass assembly of the host galaxy. Preliminary  statistical results based  on a sample of 360 massive nearby galaxies are presented. 

\keywords{Deep imaging, galaxy evolution, mass assembly}
\end{abstract}

\firstsection % if your document starts with a section,

\section{Introduction}
Several teams  are now actively involved in deep  imaging programs of nearby galaxies. These surveys  use different techniques and instruments:  long exposures either on small, amateur-like, telescopes (e.g. Martinez-Delgado et al., 2010; Rich et al., this volume) or 10-meter class ones (e.g. Trujillo \& Fliri, 2016; Tanaka et al., this volume),  instruments  dedicated to Low Surface Brightness (LSB) studies (e.g. ``Dragonfly", van Dokkum et al., 2014), shallow surveys, stacking hundreds of galaxies (e.g. SDSS, DÕSouza et al., 2014), multi-visited fields initially used for calibration purposes (SDSS-Stripe82, Fliri \& Trujillo, 2016), or  LSB optimized observations  on 2--4 meter class telescopes (among others, ESO/VST, Iodice et al., 2016; Blanco-CTIO/DECAM, Munoz et al., 2015; CFHT/MegaCam, NGVS, Ferrarese et al., 2012).  We have exploited the latter approach, and in particular MegaCam on the CFHT, as it offers a number of advantages:
(a) a (local) surface brightness limit of 29 mag.arcsec$^{-2}$  reached in only 45 min, instead of  10-30 hours when using small telescopes. This could be achieved thanks to a dedicated observing strategy and  pipeline that resulted in a gain of several mag with respect to traditional imaging surveys with the same total integration time 
(b)  the ability to conduct, though large programs,  complete volume limited surveys instead of   in-depth studies of a few specific galaxies  
(c) a good sensitivity to extended LSB structures while benefiting from the excellent image quality  of MegaCam and seeing conditions of  the Mauna Kea site. This allows us to conduct additional studies such as the census of globular clusters, another tracer of the outskirts of galaxies.

The  Large Program MATLAS, presented here, complemented by the NGVS which has mapped the whole Virgo cluster, has  targeted 240 nearby massive Early Type galaxies (ETGs)  from the Atlas3D  sample  (Cappellari et al., 2011; Duc et al., 2015). As a bonus,  120 Late Type galaxies (LTGs)  located in the large MegaCam field of view, as well as hundreds of dwarf galaxies, benefit from deep images. 

\begin{figure}[h]
% \vspace*{-2.0 cm}
\begin{center}
 \includegraphics[width=0.7\textwidth]{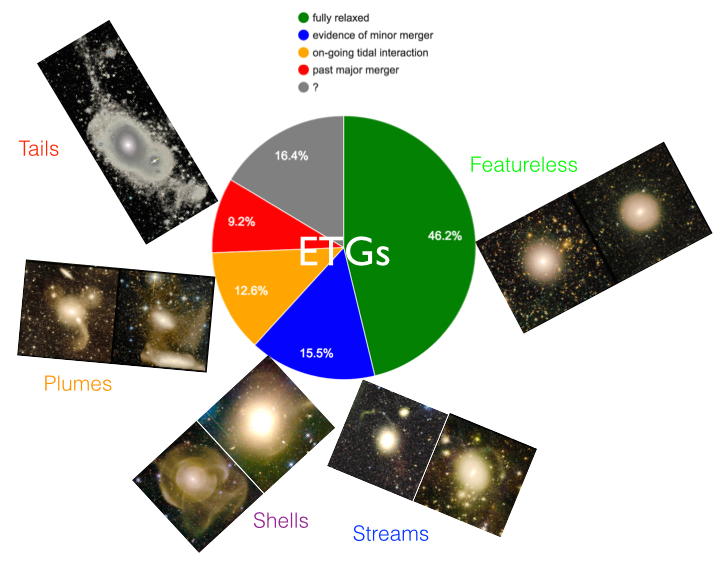} 
% \vspace*{-1.0 cm}
 \caption{The classification scheme used for the MATLAS deep imaging survey, which is based on the identification of various types of collisional debris. Examples of tails, plumes, shells and streams are illustrated in the insets. The  pie chart  summarizes the  preliminary  statistical results for the population of 240 massive Early Type Galaxies. }
\label{fig1}
\end{center}
\end{figure}

\begin{figure}[h]
% \vspace*{-2.0 cm}
\begin{center}
 \includegraphics[width=0.71\textwidth]{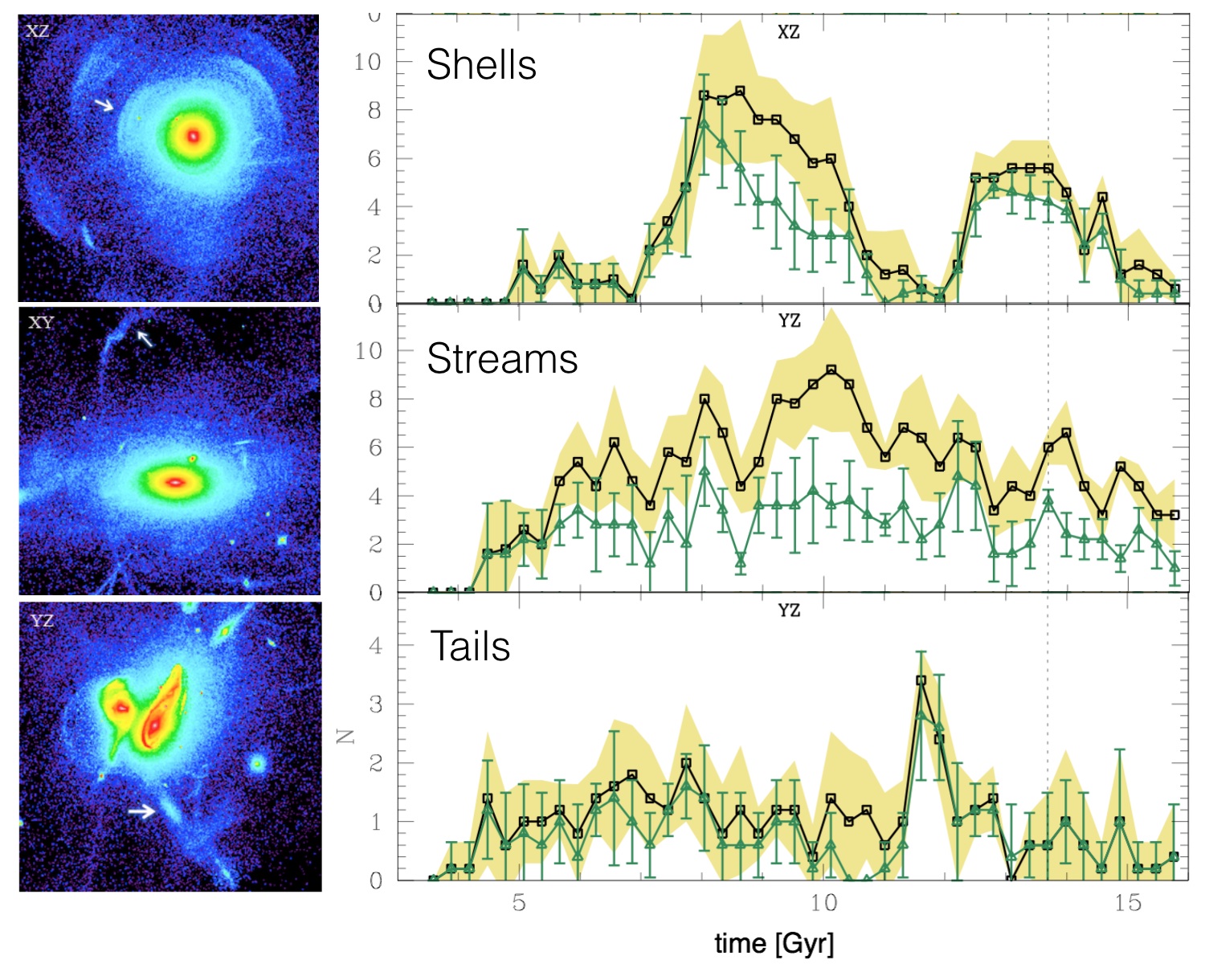} 
 \vspace*{-0.3 cm}
 \caption{Simulation of the mass assembly of a massive galaxy traced by their fine structures. The evolution of the number of shells, streams and tails as a function of time is shown to the right. The green curve relies on the analysis of snapshots cut at the limiting surface brightness of the MATLAS  observations. Examples of each of these structures are shown to the left. The images correspond to the orientation for which they are best seen. }
 \label{fig2}
\end{center}
\end{figure}

\section{Census of collisional debris in observations and simulations}
Like galaxy archeology done in local galaxies with resolved stellar populations,  the deep imaging programs  exploiting  the diffuse light  focus on the external regions of galaxies.  The individual collisional debris found there trace  the last merging events,  while the extended stellar halos have an older but less precise memory of the mass assembly. I present here our efforts to make the census of the fine structures around ETGs and LTGs, based on the CFHT multi-band deep images and the analysis of numerical simulations. 
The originality of our approach is to make a distinction between the various types of collisional debris, as each of them gives  different pieces of information on the past accretion  episodes.

As illustrated in Fig.~\ref{fig1}, minor mergers produce narrow stellar streams; major wet mergers are characterized by prominent star-forming tidal tails; major dry mergers induce the formation of faint broad gas-poor plumes; shells trace back intermediate mass mergers while the absence of any structure may indicate cold gas accretion and a secular evolution.
Streams, tails, plumes and shells were identified by eye by several team members, based on  surface brightness and color maps, as well as  residual images resulting from the subtraction of a model of the host galaxy.

Collisional debris are intrinsically faint and furthermore evaporate with time or may be destroyed by subsequent accretion events. Their detection depends on the type of structures, on their orientation on the sky, and above all on the achieved surface brightness limit. Numerical simulations are absolutely required to estimate their visibility time and thus reconstruct the past merging history of galaxies. 
  This motivated us to analyse a set of simulations made in cosmological context.  Snapshots  were produced at different times, orientations on the sky, and surface brightness limits, and then mixed. Team members were invited to identify by eye and count the various types of fine structures. The evolution of the number of structures as a function of time was then reconstructed, and compared with the merging history of the galaxy, which is known by construction (Mancillas et al., in prep, see Fig.~\ref{fig2}).

\section{Preliminary results}
Thanks to the depth of the survey, we found -- unsurprisingly -- a fraction of tidally perturbed massive galaxies raising from about 15\% in classical surveys (e.g. Atkinson et al., 2013) to 40\% (see Fig.~\ref{fig1}). Our large sample of  360 galaxies  allows us to obtain statistically significant trends.  
 In particular, ETGs with stellar masses above 10$^{11}$ M$_{\odot}$, and, in particular, among them the  slow rotators, i.e. the galaxies with a stellar kinematics showing no indication of rotation, seem to have experienced recent wet major mergers with a frequency increased by a factor of 3. While a significant fraction of LTGs in our sample show evidence for on-going tidal interactions, only few of them exhibit streams and tails indicative of a past strong  minor / major merger activity. 	
 Preliminary results further indicate a mild dependence of the fraction of tidally perturbed galaxies on the large scale environment. All these trends on the fine structure fraction, once assessed with a more detailed analysis and completed by studies of the outer stellar halo,  will  be compared with predictions from cosmological simulations.

\end{document}